\documentclass[prb,a4paper,twocolumn,floatfix,showpacs,showkeys,amsmath,amssymb,nobibnotes,altaffilletter,superscriptaddress]{revtex4-1}

\usepackage[utf8x]{inputenc}
\usepackage{graphicx}% Include figure files
\usepackage{pslatex}
\usepackage{xspace}

\newcommand{\degree}{$^{\circ}$}

\graphicspath{{./}{images/}}

\begin{document}

\title{Order of Decay of Mobile Charge Carriers in P3HT:PCBM Solar Cells}

\author{Carsten Deibel}\email{deibel@physik.uni-wuerzburg.de; deibel@disorderedmatter.eu}
\affiliation{Experimental Physics VI, Julius-Maximilian-University of Würzburg, 97074 Würzburg, Germany}
\author{Daniel Rauh}
\affiliation{Bavarian Center for Applied Energy Research e.V. (ZAE Bayern), D-97074 W\"urzburg, Germany}
\author{Alexander Foertig}
\affiliation{Experimental Physics VI, Julius-Maximilian-University of Würzburg, 97074 Würzburg, Germany}

\date{\today}

\begin{abstract}
The charge carrier dynamics of organic solar cells are strongly influenced by trapping and allow to draw conclusions on the loss mechanisms limiting the photovoltaic performance. In this study we derive the recombination order $\Delta$ of mobile charge carriers. For annealed P3HT:PCBM solar cells, it allows us to pinpoint the dominant recombination of mobile with trapped charge carriers in tail states. While the characteristic tail state energy of about 40~meV rises to about 100~meV for 30~h oxygen exposure under illumination, $\Delta$ decreases only weakly from 1.70 to 1.62: This corresponds to a slight shift towards trap-assisted recombination.
\end{abstract}

\pacs{}

\keywords{organic solar cells, charge carrier recombination, trap states, photocurrent}

\maketitle

%\section{Introduction}

The nongeminate recombination of charge carriers is the dominant loss mechanism in state-of-the-art organic solar cells.\cite{deibel2010review} The corresponding dynamics of the charge carrier concentration $n$ are often observed as power law decay $n(t) \propto t^{-(\delta-1)}$ by transient absorption\cite{montanari2002,offermans2003} and charge extraction techniques.\cite{shuttle2008,deibel2008,juska2008,street2011,mauer2011,kniepert2011,mackenzie2012,rauh2012} The loss currents $j_{loss}$ and recombination rates $R$
\begin{equation}
	j_{loss} \propto R \propto n^\delta 
	\label{eqn:delta}
\end{equation}
then have orders of decay $\delta$ which exceed the value of two expected for recombination of electrons and holes in a homogeneous system without trapping. An important question is, how relevant is the recombination order from transient measurements for the device performance under steady state operating conditions? Our aim is to answer this question by relating $\delta$ to another figure of merit. 

The photocurrent of a solar cell is maintained only by transport of the mobile fraction of the excess charge carriers, as trapped charge carriers are immobile.\cite{rudenko1982,baranovskii2000} Therefore, in order to judge the impact of loss mechanisms on the photocurrent, it is instructive to define the order of decay $\Delta$ of the density of \emph{mobile} charge carriers $n_c$. Then,
 \begin{equation}
	j_{loss} \propto n_c^{\Delta} .
	\label{eqn:Delta}
\end{equation}
Here, $n=n_c+n_t$, where $n_t$ is the trapped charge carrier density. For recombination of only mobile charge carriers, $\Delta=2$ is expected, whereas trap-assisted recombination yields a value of $1$. For values inbetween, it signifies the relative dominance of these two mechanism.

%\begin{figure}[tb] 
%	\centering
%	\includegraphics[width=6.75cm]{iv+params.pdf}
%	\caption{Solar cell parameters normalised to the values of the fresh P3HT:PCBM solar cells. The loss of $PCE$ upon oxygen exposure is mainly due to loss of short circuit current, whereas open circuit voltage and fill factor remain almost constant.}
%	\label{fig:iv+params}
%\end{figure}

With this paper, we offer a derivation of the recombination order for mobile charge carriers, $\Delta$. We explain its importance for characterising the dominant charge carrier loss mechanism, and relate it to the less transparent order of decay $\delta$. For annealed P3HT:PCBM solar cells, we find that recombination of mobile with trapped charge carriers are dominant, and find a characteristic (exponential) tail state energy of about 40~meV in agreement with literature. Exposure to oxygen in dark and under illumination for up to 30~h leads to a significant increase of the characteristic energy to about 100~meV. Our analysis of $\Delta$ allows us to conclude that this change signifies only a slight shift towards more trap-assisted recombination.

%\section{Experimental}

To prepare the solar cells, a 40~nm thick layer of poly(3,4-ethylenedioxythiophene):poly(styrenesulfonate) (CLEVIOS PVP AI4083) was spincoated on top of indium tin oxide/glass substrates. The substrates were transferred into a nitrogen glovebox, then annealed at 130\degree{}C for 10~min. The $L=190$~nm thick active layer consisting of poly(3-hexylthiophene-2,5-diyl) (P3HT):[6,6]-phenyl-C$_{61}$ butyric acid methyl ester (PCBM) blend (weight ratio of 1:0.8) was spin coated from chlorobenzene solution and thermally annealed at 130\degree{}C for 10~min subsequently. As cathode a Ca (3~nm)/Al (90~nm) electrode was thermally evaporated, with an active area of about 9~mm$^2$. % 9.32~mm$^2$. 

After current--voltage (IV) characterisation (Keithley 2602) under AM1.5g simulated illumination the samples were transferred to a Janis CCS 550 He contact gas cryostate, being exposed to ambient air for a few minutes. The illumination was provided by a high power light emitting diode (LED) with 10~W electrical power (Seoul P7 Emitter). The calibration of the LED illumination level was performed using a silicon solar cell.% with a linear response of the short circuit current on the incident light intensity.

Charge extraction measurements were performed using the LED and a double pulse generator (Agilent 81150A) for applying the premeasured open circuit voltage $V_{oc}$ to the solar cell. At a certain time $t_0$, the LED was switched off by shorting the constant current source (Keithley 2602) with a high power transistor triggered by the double pulse generator. The resulting current was preamplified by a FEMTO DHPCA-100 current--voltage amplifier and recorded by an Agilent DSO 90254A oscilloscope. The time integration of the current was corrected for charges stored on the electrodes, yielding the desired charge carrier densities.

During the measurements the devices were stored in He atmosphere, during the degradation steps the cryostate was filled with synthetic air (80~\%~N$_2$, 20~\%~O$_2$, no moisture). The measurements were done before starting the degradation, after 1, 3, 10 and 30~h in dark. Afterwards the degradation under 1~sun illumination was started and the cell was characterized again after 1, 3, 10 and 30~h.

%\section{Experimental Results}

%\subsection{Steady-State Measurements}

The P3HT:PCBM device used for this study showed solar cell parameters of open circuit voltage $V_{oc}=0.57$~V, short circuit current density $j_{sc}=8.2$~mA/cm$^2$, fill factor $FF=69$~\%, yielding a power conversion efficiency $PCE=3.2$~\%. For dark oxygen exposure only $j_{sc}$ dropped strongly by about 30~\%, whereas $V_{oc}$ remained constant and the fill factor increased by 3~\%. For the additional oxygen exposure under illumination $j_{sc}$ continued to decrease to about 60~\% of its initial value, $FF$ decreased back to 100~\% and $V_{oc}$ to 97~\% of their initial values after additional 30~h under illumination, respectively.% (Fig.~\ref{fig:iv+params}).
These changes are consistent with Ref.~\onlinecite{schafferhans2010}, although there they decreased more quickly, which we assign to the low UV emission of the LED used in the present study. 

\begin{figure}[tb] 
	\centering
	\includegraphics[width=7.5cm]{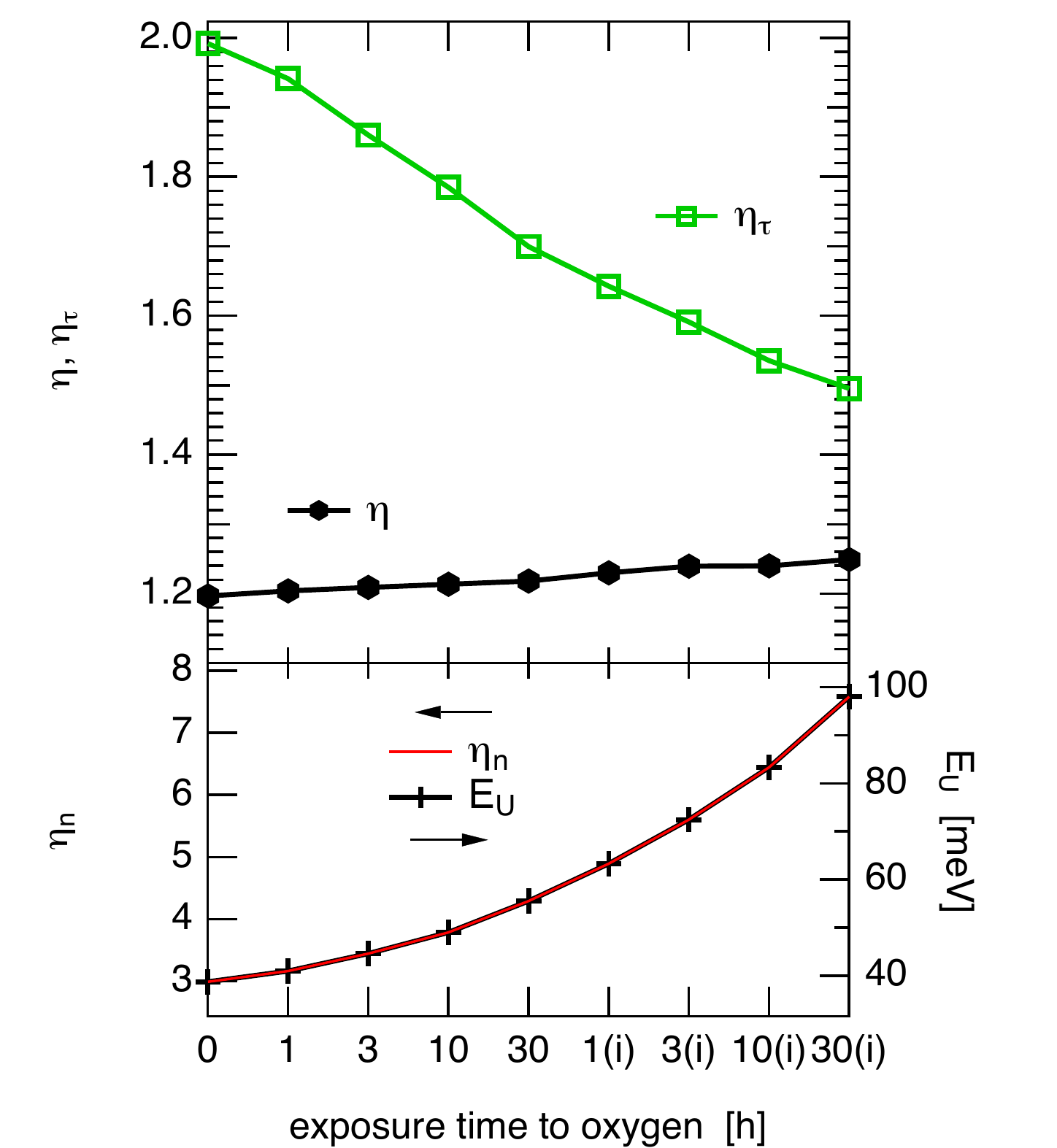}
	\caption{Change of the diode ideality factor $\eta$ and the ideality factors for the charge carrier concentration $\eta_n$ (charge extraction data and Eqn.~(\ref{eqn:n})) and lifetime $\eta_\tau$ (Eqn.~(\ref{eqn:ideality})). The characteristic tail energy $E_U$, calculated from $\eta_n$ by Eqn.~(\ref{eqn:Eu}), strongly increases with oxygen exposure time.}
	\label{fig:ideality_and_tail-energy}
\end{figure}

The diode ideality factor $\eta$ determines the voltage dependence of the diode current and provides information on the dominant recombination mechanism.\cite{shockley1952,kirchartz2011} It can be extracted using the Shockley diode equation under illumination, as
\begin{equation}
	\eta = \left(\frac{kT}{q}\right)^{-1}\frac{dV_{oc}}{d\ln(j_{sc})} ,
\end{equation}
if the photogeneration is field independent. $q$ is the elementary charge, $kT$ the thermal energy, We determined $\eta$ in the range from 0.001 to 0.4~suns. As shown in Fig.~\ref{fig:ideality_and_tail-energy} (top), it increases continuously with oxygen exposure time. % from an initial value of 1.20 to 1.22 after 30~h of dark degradation and to 1.25 after the additional 30~h exposure under illumination. 

\begin{figure}[tb] 
   \centering
    \includegraphics[width=7cm]{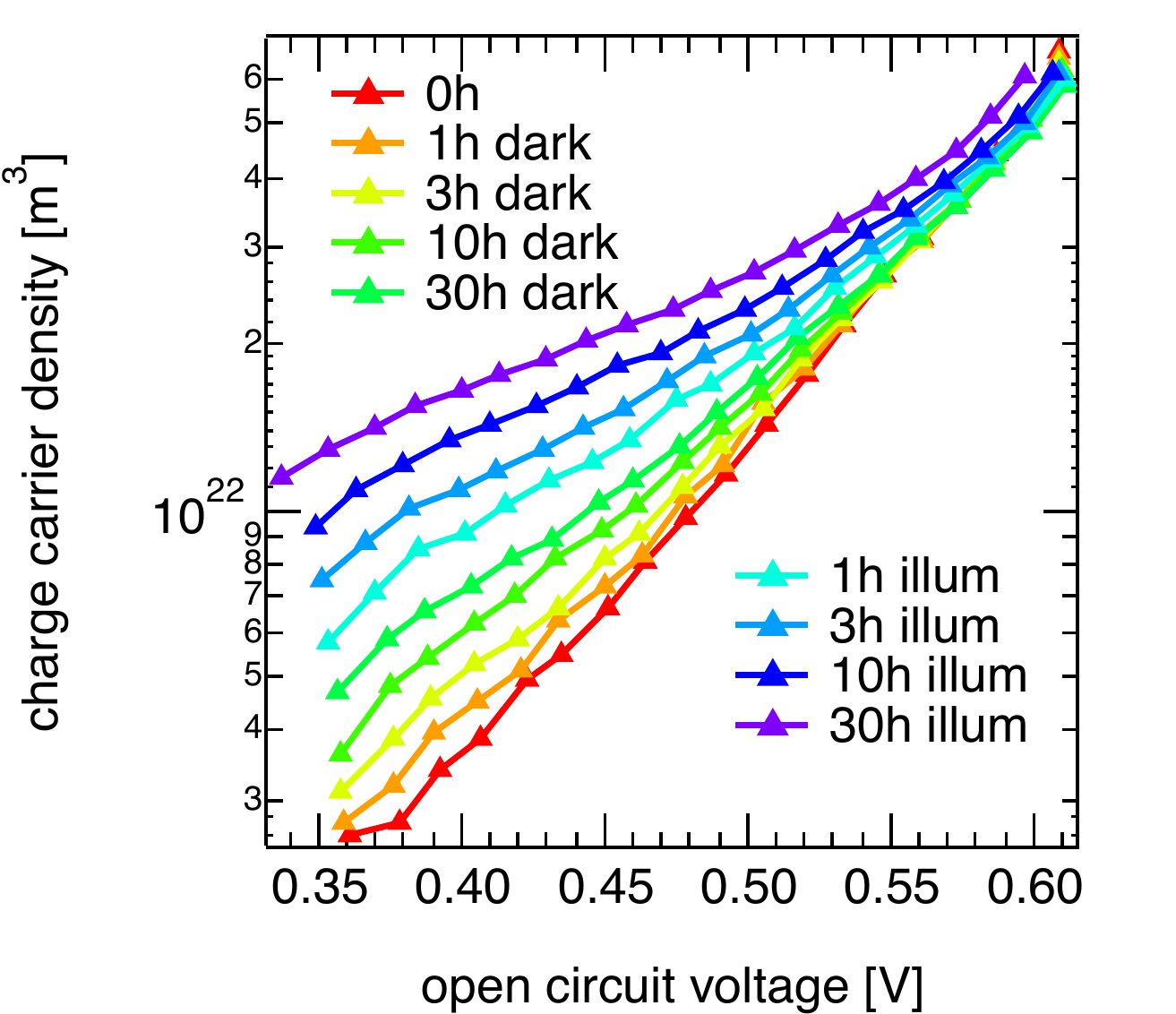} 
    \caption{The charge carrier density at open circuit conditions as obtained from charge extraction measurements for every degradation step.}
     \label{fig:n_voc_deg}   
\end{figure}

The charge carrier density $n$ from charge extraction measurements at open circuit conditions is shown in Fig.~\ref{fig:n_voc_deg}. For the nonexposed device, $n$ shows the typical exponential dependence on $V_{oc}$,\cite{kirchartz2011,foertig2012}
\begin{align}
	n = n_0 \exp{ \left( \frac{q V_{oc}}{\eta_n kT} \right) } ,
	\label{eqn:n}
\end{align}
where $n_0$ is the dark carrier concentration and $\eta_n$ the ideality factor for the carrier concentration.\cite{foertig2012} Here and in the following, we assume that deeply trapped charge carriers, which cannot be extracted, do not play a major role in the recombination process.
$\eta_n$ signifies the voltage dependence of $n$ and rises strongly with oxygen exposure time (Fig.~\ref{fig:ideality_and_tail-energy} (bottom)). For mobile charge carriers, $\eta_n=2$ would be expected, as their concentration is proportional to $\exp(qV/2kT)$ (c.f.~Eqn.~(\ref{eqn:n})).\cite{kirchartz2011} 

In the same way, $\eta_\tau$ determines the voltage dependence of the effective charge carrier lifetime $\tau$, 
\begin{equation}
	\tau \propto \exp \left( -\frac{qV_{oc}}{\eta_\tau kT} \right) .
	\label{eqn:tau}
\end{equation}
By using\cite{foertig2012}
\begin{equation}
	\eta^{-1} = \eta_n^{-1} + \eta_\tau^{-1} ,
	\label{eqn:ideality}
\end{equation}
we are able to calculate $\eta_\tau$ from the experimentally determined $\eta$ and $\eta_n$. Ideally, a value of $\eta_\tau$ of 2 is expected,\cite{kirchartz2011} and this is indeed the case for the fresh device. Recently it was shown that $\eta_\tau<2$ can be due to inhomogeneous spatial charge carrier distributions in the device,\cite{kirchartz2012} but clearly in our thick device it is rather homogeneous. However, we see in Fig.~\ref{fig:ideality_and_tail-energy} (top) that $\eta_\tau$ decreases notably upon oxygen exposure down to about 1.5 for 30~h illumination, which is consistent with measurements of the transient photovoltage we performed on similar samples.

%In the following, our aim is to investigate the nongeminate loss mechanism in P3HT:PCBM solar cells. 

%\section{Discussion}	

%\subsection{Recombination Mechanism and Order of Decay}

The nongeminate recombination rate $R$ is empirically characterised by the order of decay $\delta$, c.f.\ Eqn.~(\ref{eqn:delta}). A $\delta$ of 2 is expected in homogeneous systems without trapping, $R\propto np = n^2$. In disordered polymer--fullerene solar cells, trapping in tail states has been observed experimentally.\cite{vandewal2009,schafferhans2008,schafferhans2010,presselt2012} Then, the recombination rate in a homogeneous system becomes
\begin{equation}
	R \propto n_c p_c + n_c p_t + n_t p_c \propto np - n_t p_t,
	\label{eqn:Rcomplete}
\end{equation}
with $n_c \ll n_t$, $p_c \ll p_t$. Here we have neglected that these contributions to recombination may have different prefactors. Clearly, not all charge carriers are available for recombination at a given time, i.e., only the mobile ones can actively meet their opposite (mobile or trapped) counter charges to drive recombination. We propose that this mechanism is the main origin of the \emph{reduced} Langevin recombination.\cite{pivrikas2005,deibel2010review} The loss of the overall charge carrier concentration is lowered, and the order of decay $\delta$ increases beyond the value of two.

Annealed P3HT:PCBM solar cells at room temperature have recently been shown to behave accordingly, with a dominant recombination of mobile with shallow trapped charge carriers.\cite{kirchartz2011,street2011,foertig2012} The contributions of electrons and holes cannot be directly distinguished, therefore we simplify Eqn.~(\ref{eqn:Rcomplete}) to
\begin{equation}
	R \propto n_c (n_c + n_t) \approx n_c n_t .
	\label{eqn:Rct}
\end{equation}
The characteristic energy of the exponential tails of the density of states distribution was determined using $\delta$,\cite{kirchartz2011} a method which was later refined to use $\eta_n$,\cite{foertig2012} 
\begin{equation}
 	E_U = \eta_n \frac{kT}{2} .
	\label{eqn:Eu}
\end{equation}
We find that $E_U$ strongly increases with oxygen exposure time from the initial 40~meV---consistent with literature\cite{kirchartz2011,mackenzie2011}---to about 100~meV, as shown in Fig.~\ref{fig:ideality_and_tail-energy} (bottom). We point out (a) that exponential tails are only approximations,\cite{foertig2012,mackenzie2012}, and (b) that we cannot distinguish between electrons and holes with charge extraction measurements, even though the trap density of states may only stem from one of the two blend constituents. 
%Also, (c) the value of $E_U$ could not be accurately determined anymore if recombination occured not only between mobile and shallowly trapped carriers, but in part through deep recombination centers. 
%Using defect spectroscopy, we determined a somewhat higher characteristic tail state energy of 57~meV for an annealed P3HT--PCBM blend.\cite{foertig2012}

\begin{figure}[tb] 
	\centering
	\includegraphics[width=7cm]{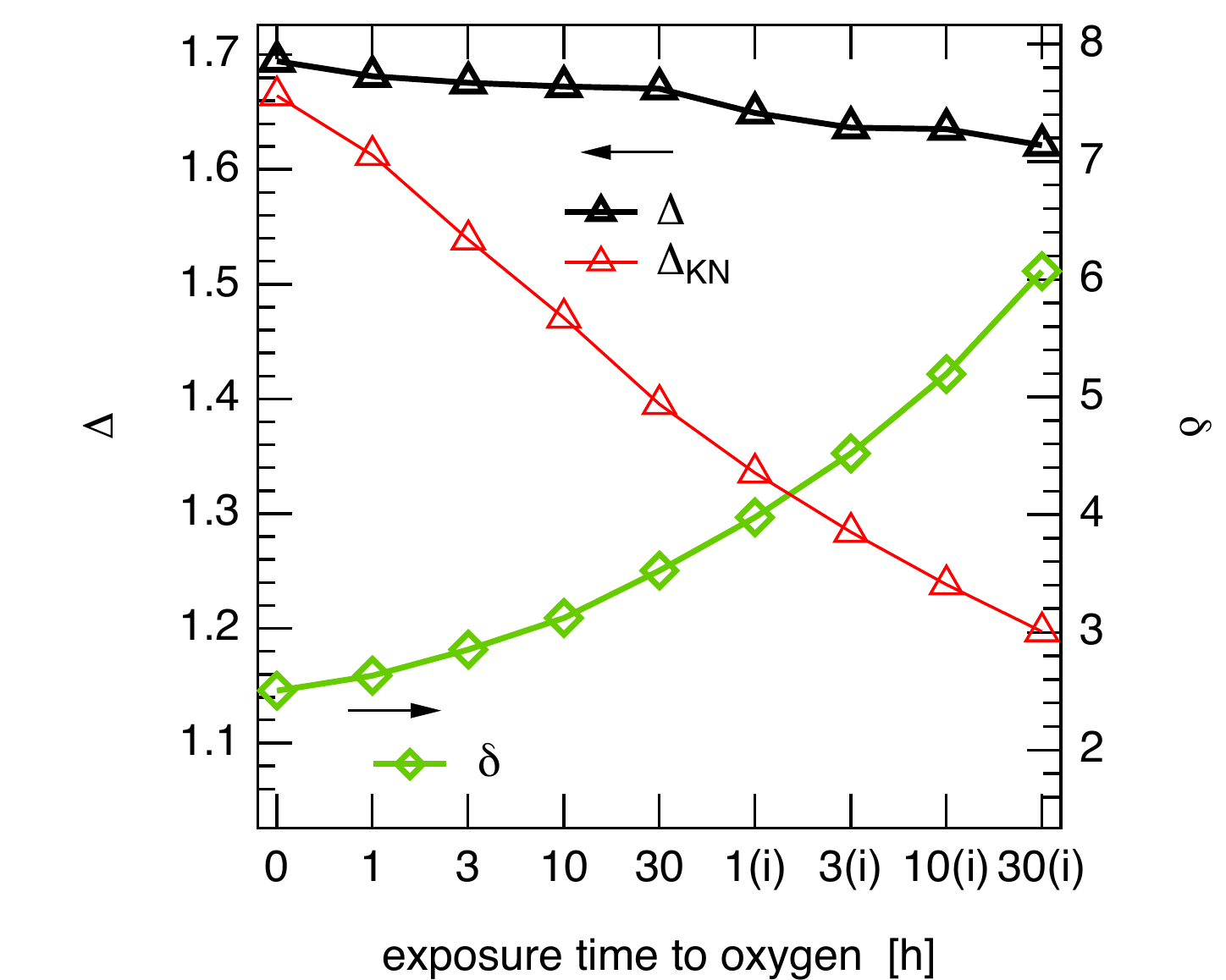}
	\caption{The recombination order $\delta$ of all charge carriers (Eqn.~(\ref{eqn:lambda})) strongly increases with oxygen exposure time, in contrast the recombination order $\Delta$ of mobile charge carriers (Eqn.~(\ref{eqn:Delta})) decreases slightly. $\Delta_{KN}$, defined in Ref.~\onlinecite{kirchartz2012}, overestimates this trend.}
	\label{fig:rec-order}
\end{figure}

The corresponding recombination order (Eqn.~(\ref{eqn:delta})) can be calculated as\cite{foertig2012} 
\begin{equation}
	\delta = \frac{\eta_n}{\eta_\tau} + 1 .
	\label{eqn:lambda}
\end{equation}
Eqn.~(\ref{eqn:lambda}) allows to obtain information on the decay dynamics of the charge carrier concentration, as shown in Fig.~\ref{fig:rec-order}: $\delta$ increases strongly with oxygen exposure time.
%, to which $\eta_n$ contributes about $3/4$ and $\eta_\tau$ only to about $1/4$. 
Clearly, $\delta > 2$, which implies that charge carrier doping or trapping are involved, and that the overall concentration $n$ of mobile and shallow trapped charge carriers decays more slowly with time. However, it remains unclear how much impact such a strong increase in characteristic tail state energy and order of decay may have on the organic solar cell under operating conditions. To find out, we consider the recombination order of mobile charge carriers.

% The contribution of $\eta_\tau$ is briefly discussed in Appendix~\ref{appendix:k-vs-etatau}. 

%\subsection{Recombination Order of Mobile Charge Carriers}

A recombination order referring only to the mobile charge carriers $n_c$ may be a relevant figure of merit, as the (photo)current is solely due to the transport of mobile charge carriers.\cite{monroe1985,baranovskii2000} $\Delta$ is defined by Eqn.~(\ref{eqn:Delta}) and was given by Kirchartz and Nelson\cite{kirchartz2012} as \begin{equation}
	\Delta_{KN} = (\delta-1)^{-1}+1
	\label{eqn:Delta_KN}
\end{equation}
for the common case of mobile-to-trapped charge carrier recombination. Within our framework, we derive a more general result for the same case, i.e.\ recombination rates following Eqn.~(\ref{eqn:Rct}):
\begin{align}
	R & \propto n^\delta \approx n_t^\delta \qquad \text{ (for } n_c \ll n_t \approx n \text{)}\\ 
	  & = n_c^{\frac{2}{\eta_n}\delta}  = n_c^{\frac{2}{\eta}} \equiv n_c^\Delta .
	\label{eqn:R-delta}
\end{align}
Here, we used $n_t \propto n_c^\frac{2}{\eta_n}$,\cite{foertig2012} and $\eta = \eta_n/\delta$ from Eqns.~(\ref{eqn:ideality}) and~(\ref{eqn:lambda}).
The resulting recombination order
\begin{equation}
	\Delta =\frac{2}{\eta} 
	\label{eqn:Delta_xxx}
\end{equation} 
is equivalent to Eqn.~(\ref{eqn:Delta_KN}) only in the case of $\eta_\tau=2$, a condition which does not necessarily hold: for instance, a field dependent mobility---which has been described by simulations\cite{koster2010} and found experimentally\cite{albrecht2012a}---will lead to $\eta_\tau\not=2$.

From Eqn.~(\ref{eqn:R-delta}), the connection between $\Delta$ and $\delta$ can be directly evaluated to $\Delta = \frac{2}{\eta_n} \delta$. With this relation, it is now possible to transform the intransparent recombination order $\delta$, describing the power law decays often found in charge carrier decay dynamics, into the figure of merit $\Delta$, which bears direct relevance to the organic solar cell performance under steady state conditions. This recombination order of mobile charge carriers allows us to directly determine the dominant nongeminate loss mechanism. A value of $\Delta=1$ corresponds to a first order charge carrier decay, for instance due to recombination exclusively through recombination centers. In contrast, $\Delta=2$ implies that all mobile charge carriers can recombine with one another. While the meaning of the ideality factor in terms of recombination has been long known, we provide here a direct, quantitative relation to the recombination order for mobile charge carriers \emph{and} the overall recombination order. A high order of decay clearly exceeding the value of two, as e.g. determined by transient absorption, can now be translated into a measure relevant for the device performance of organic solar cells.

In Fig.~\ref{fig:rec-order}, we assembled the resulting orders of decay for all and for mobile charge carriers, $\delta$ and $\Delta$, respectively. While $\delta$ increases strongly upon oxygen exposure, as discussed above, the order of decay for mobile charge carriers decreases from 1.70 to 1.62 slightly towards more trap-assisted recombination. In contrast, $\Delta_{KN}$ calculated according to Kirchartz and Nelson\cite{kirchartz2012} decreases more strongly, potentially leading to an overestimation of the importance of first order recombination.

%\section{Conclusions}

In conclusion, we found the recombination order of the mobile fraction of charge carriers, $\Delta$, to be twice the inverse diode ideality factor. This figure of merit allows to evaluate the dominant nongeminate losses of organic solar cells relevant for operating conditions. For annealed P3HT:PCBM solar cells we found the recombination of mobile and trapped charge carriers to be dominant. Exposure to oxygen, partly under illumination, lead to an increase of the characteristic tail energy from about 40~meV to 100~meV for 30~h oxygen exposure under illumination. Despite this strong change, the recombination order of mobile charge carriers $\Delta$ is reduced slightly from 1.70 to 1.62 due to oxygen exposure, representing only a weak shift further towards trap-assisted recombination. With our approach we can show that despite the strong increase in characteristic tail state energy and order of decay, the organic solar cell under operating conditions is not strongly impacted by \emph{nongeminate} recombination of charge carriers. Thus, $\Delta$ is a useful figure of merit to directly determine the recombination order relevant to the photocurrent from the diode ideality factor.

\begin{acknowledgments}
The authors thank Vladimir Dyakonov, Alexander Wagenpfahl, Andreas Baumann and Jens Lorrmann for interesting discussions. The current work was supported by the Deutsche Forschungsgemeinschaft within the PHORCE project (Contract No.~DE~830/8-1). C.D. gratefully acknowledges the support of the Bavarian Academy of Sciences and Humanities.
\end{acknowledgments}

\end{document}